\documentclass{PoS}
\usepackage{epsfig}
\title{HERAPDF fits including $F_2$(charm) data}

\ShortTitle{HERAPDF fits including $F_2$(charm) data}

\author{\speaker{A. Cooper-Sarkar}\thanks{On behalf of H1 and ZEUS Collaborations}\\
        Oxford University, UK\\
        E-mail: \email{a.cooper-sarkar@physics.ox.ac.uk}}

\abstract{PDF fits in the HERAPDF1.0 formalism have been made to the combined 
HERA-I inclusive data and the combined $F_2$(charm) data from the H1 and 
ZEUS experiments. The charm data are 
found to be sensitive to the value of the charm mass and the choice of the 
heavy quark scheme. This has consequences for the predictions of $W$ and $Z$
cross-sections at the LHC. 
         }

\FullConference{XVIII International Workshop on Deep-Inelastic Scattering and Related Subjects, DIS 2010\\
		April 19-23, 2010\\
		Firenze, Italy}

\begin{document}

\section{Introduction}
The HERAPDF1.0 set of parton density functions was extracted from a fit to the 
combined inclusive deep-inelastic scattering data from the H1 and ZEUS 
experiments~\cite{herapdf10}. The predictions of these PDFs give a good 
description of the 
newly combined HERA $F_2$(charm) data, for $Q^2 > 4~$GeV$^2$. 
This PDF analysis used a central value for the charm quark mass 
$m_c=1.4~$GeV following previous PDF analyses~\cite{mstwcteq}. However a model 
uncertainty, $1.35 < m_c < 1.65$~GeV, was also evaluated. The newly combined 
charm data are 
sensitive to the choice of charm mass, thus a reduction in model uncertainty 
should be possible However, the HERAPDF1.0 used a 
specific heavy quark mass scheme - the general-mass variable-flavour-number 
scheme of Thorne and Roberts (RT-VFN)~\cite{trvfn}. The present contribution 
describes 
fits made in the HERAPDF formalism, including the combined HERA $F_2$(charm) 
data, using various different heavy quark mass schemes and various values 
of the charm quark mass. Full details are given in the orginal 
talk~\cite{mytalk}. 

\section{HERAPDF Fits including $F_2$(charm) data}
\begin{figure}[tbp]
%\vspace{-1.0cm} 
%\vspace*{5pt}
\centerline{
\epsfig{figure=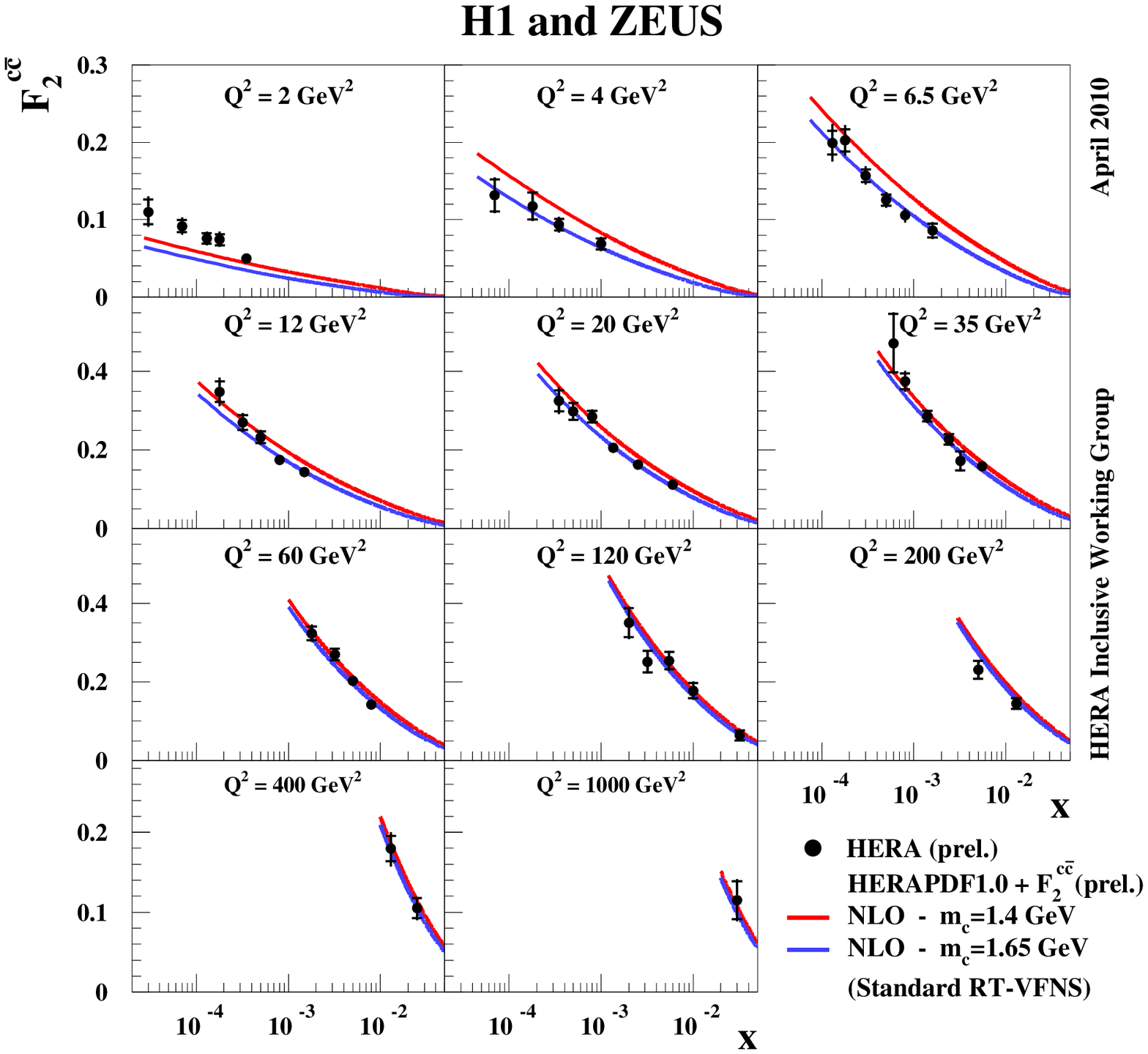,width=0.5\textwidth,height=6.0cm}
\epsfig{figure=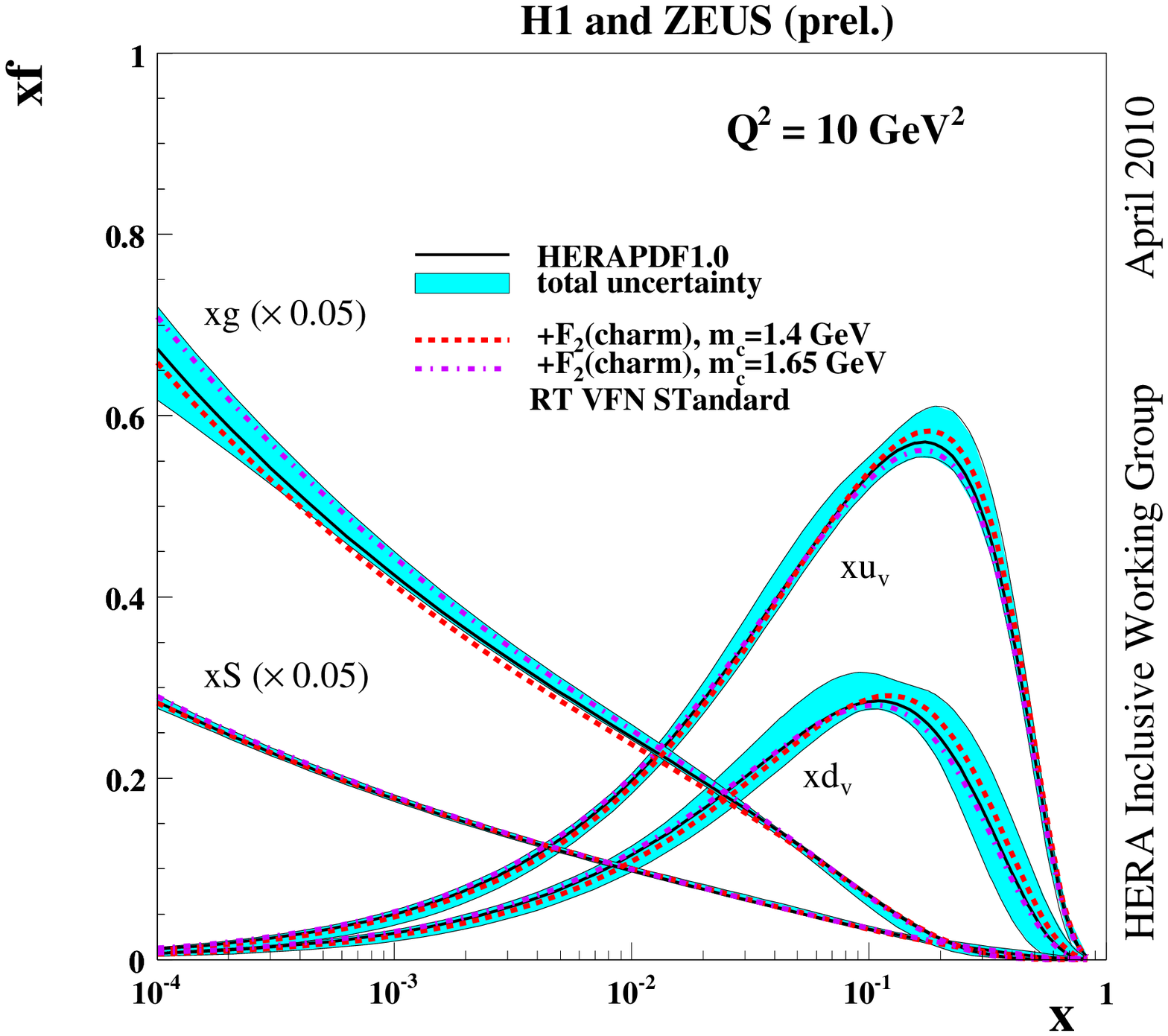,width=0.5\textwidth,height=6.0cm}}
\caption{Left: comparison of $F_2$(charm) data to a PDF fit which includes 
these data for $m_c=1.4$ and $1.65~$GeV, using the standard RT-VFN scheme. 
Right: the PDFs which correspond to these two fits.}
\label{fig:rtvfndatpdf}
\end{figure}

Firstly the HERA combined $F_2$(charm) data are included in the fit 
using the standard version of the RT-VFN scheme as used for MSTW08 PDFs. 
The usual cut, 
$Q^2 > 3.5~$GeV$^2$, is applied to these fits such that there are 
41 charm data 
points in addition to the 592 data points from combined HERA-I inclusive data 
on Neutral Current (NC) and Charged Current (CC) $e^+p$ and $e^-p$ scattering.
Charm mass values, $m_c=1.4~$GeV, and the pole-mass, $m_c=1.65~$GeV, 
are investigated.
The charm data prefer the higher value, see Table~\ref{tab:chisq}. These fits
and the PDFs which correspond to them are shown in Fig.~\ref{fig:rtvfndatpdf}.
The larger charm mass suppresses charm 
production and the gluon PDF which corresponds to it is enhanced at low $x$.
\begin{table}[h]
\centerline{
\begin{tabular}{|l|l|r|}
%\vspace{-1.0cm}
\hline
 Scheme&  total $\chi^2$/ndp & $F_2$(charm) $\chi^2$/ndp  \\
\hline
RTVFN Standard ($m_c=1.4$) & 730.7/633  & 134.5/41    \\
RTVFN Standard ($m_c=1.65$) & 627.5/633  & 43.5/41    \\
RTVFN Optimized ($m_c=1.4$) & 644.6/633  & 64.8/41    \\
RTVFN Optimized ($m_c=1.65$) & 695.4/633  & 100.1/41    \\
ACOT ($m_c=1.4$) & 644.6/633  & 89.5/41    \\
ACOT ($m_c=1.65$) & 605.7/633  & 41.4/41    \\
FFN ($m_c=1.4$) & 567.0/565  & 51.7/41    \\
FFN ($m_c=1.65$) & 852.0/565  & 248.9/41    \\
NNLO ($\alpha_s=0.1176$, $m_c=1.4$) & 703.1/633  & 60.3/41    \\
NNLO ($\alpha_s=0.1176$, $m_c=1.65$) & 832.9/633  & 185.7/41    \\
NNLO ($\alpha_s=0.1145$, $m_c=1.4$) & 681.1/633  & 54.5/41    \\
NNLO ($\alpha_s=0.1145$, $m_c=1.65$) & 862.3/633  & 198.0/41    \\
\hline
\end{tabular}}
\caption{$\chi^2$ per data point for HERAPDF fits to HERA-I data and combined 
$F_2$(charm) data for various values of the charm mass and various heavy 
quark schemes. Fits are made at NLO unless otherwise stated.
}
\label{tab:chisq}
\end{table}

\begin{figure}[tbp]
%\vspace{-1.0cm} 
%\vspace*{5pt}
\centerline{
\epsfig{figure=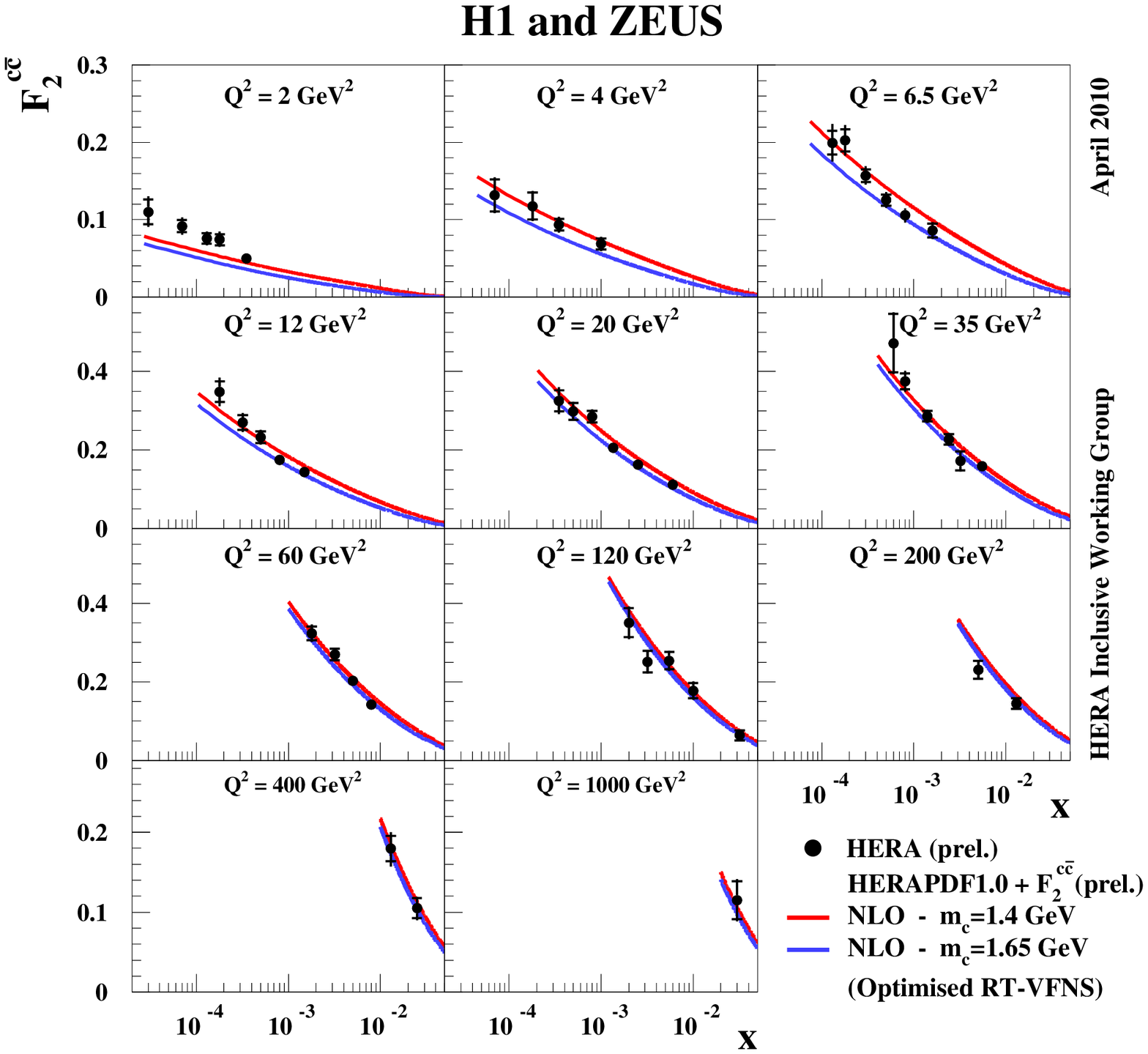,width=0.5\textwidth,height=6.0cm}
\epsfig{figure=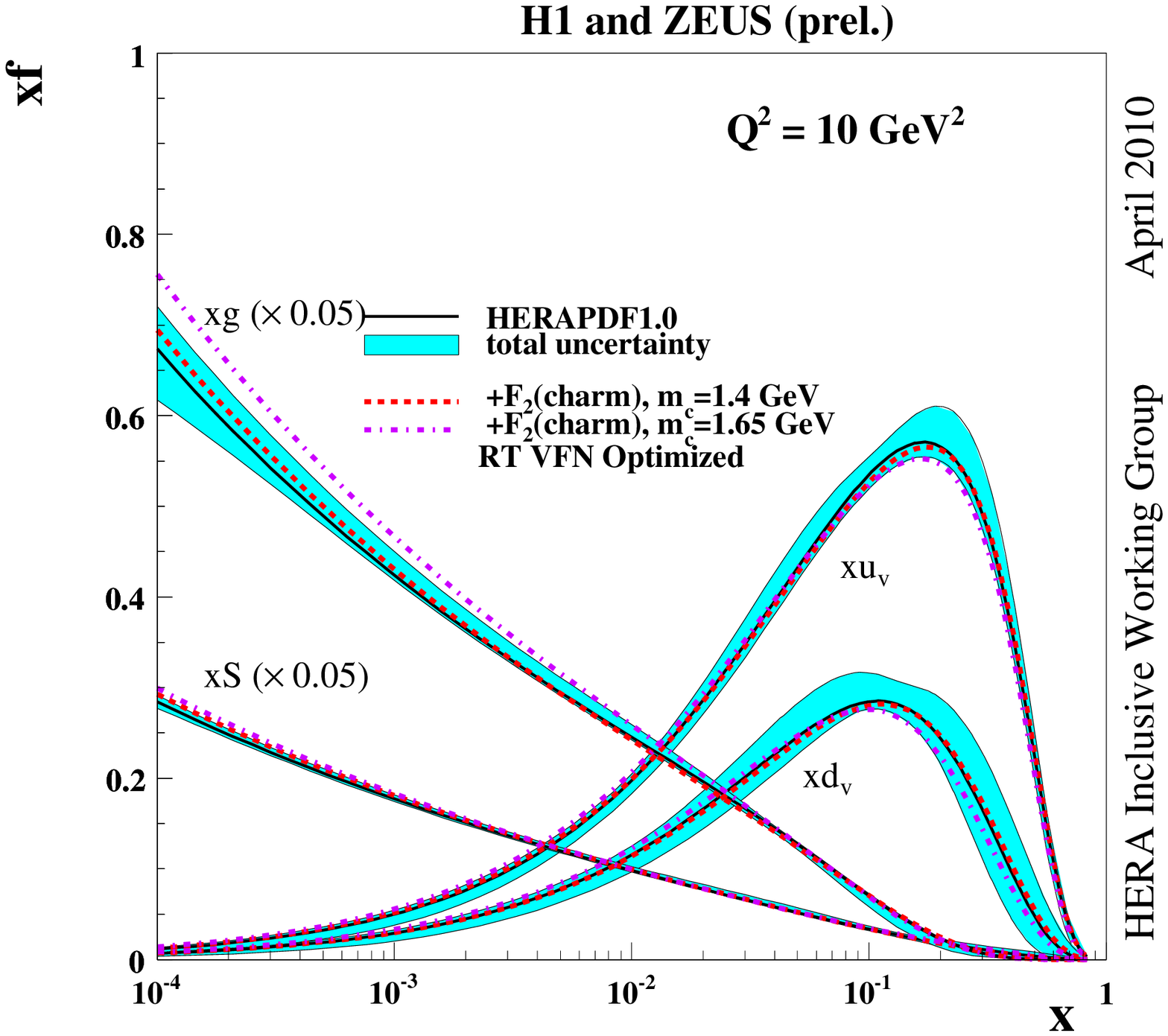,width=0.5\textwidth,height=6.0cm}}
\caption {Left: comparison of $F_2$(charm) data to a PDF fit which includes 
these data for $m_c=1.4$ and $1.65~$GeV, using the optimized RT-VFN scheme. 
Right: the PDFs which correspond to these two fits.}
\label{fig:rtvfnoptdatpdf}
\end{figure}

Thorne has suggested possible modifications of the heavy quark 
scheme~\cite{Thorne}. An optimized scheme has been selected for study 
because of its smooth threshold behaviour. In this scheme the charm data prefer
mass charm, $m_c=1.4~$GeV, see Table~\ref{tab:chisq}. 
The fits to the data and the resulting PDFs are  
shown in Fig.~\ref{fig:rtvfnoptdatpdf}. The smoother threshold 
behaviour of this scheme can be clearly seen. Charm production at threshold is 
somewhat suppressed in this scheme relative to the standard scheme so that a 
large charm mass is not needed for a good fit to data. The gluon PDF is 
somewhat enhanced at low $x$ in the optimal scheme as compared to the 
standard scheme. 

\begin{figure}[tbp]
%\vspace{-1.0cm} 
%\vspace*{5pt}
\centerline{
\epsfig{figure=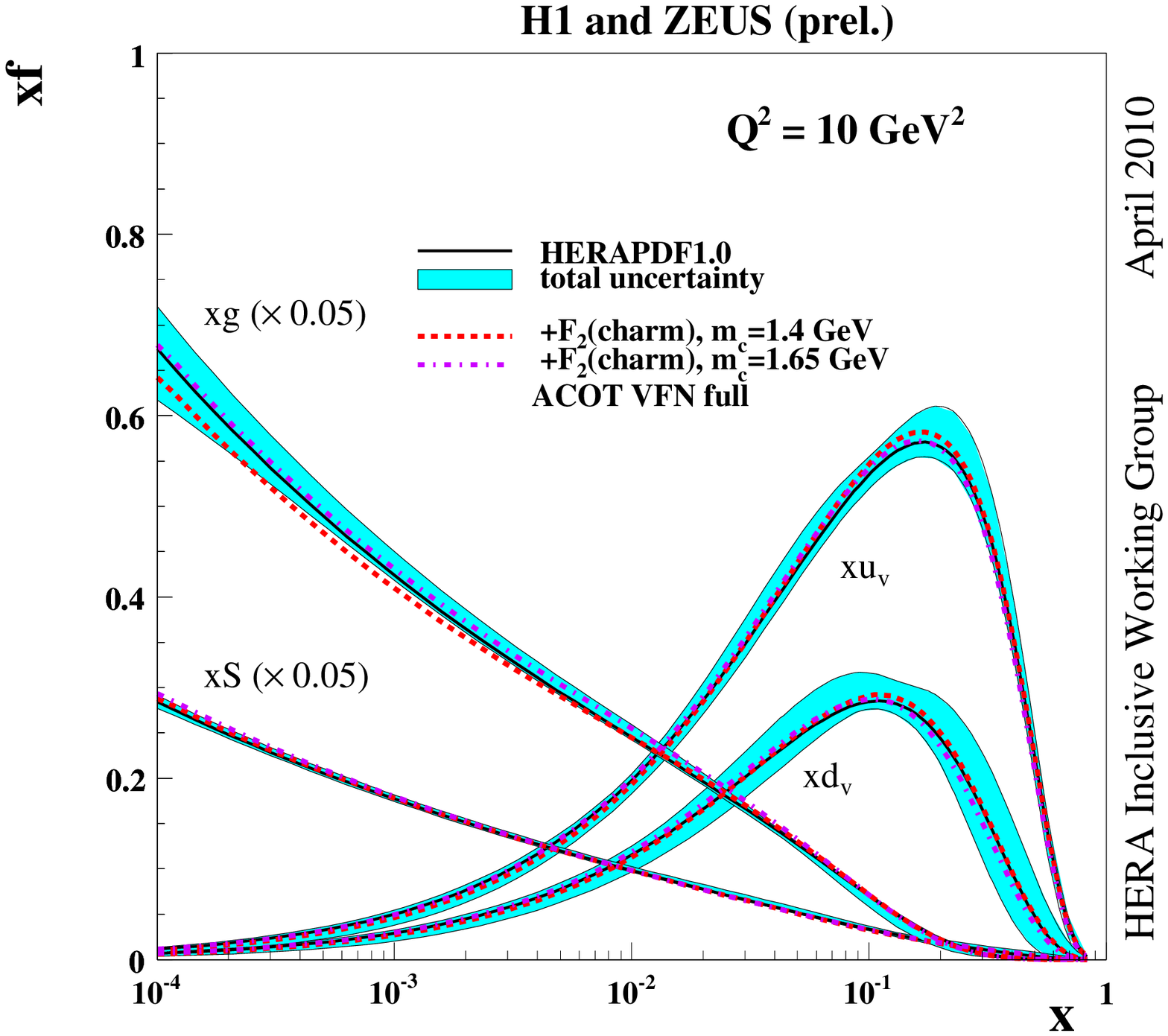,width=0.5\textwidth,height=6.0cm}
\epsfig{figure=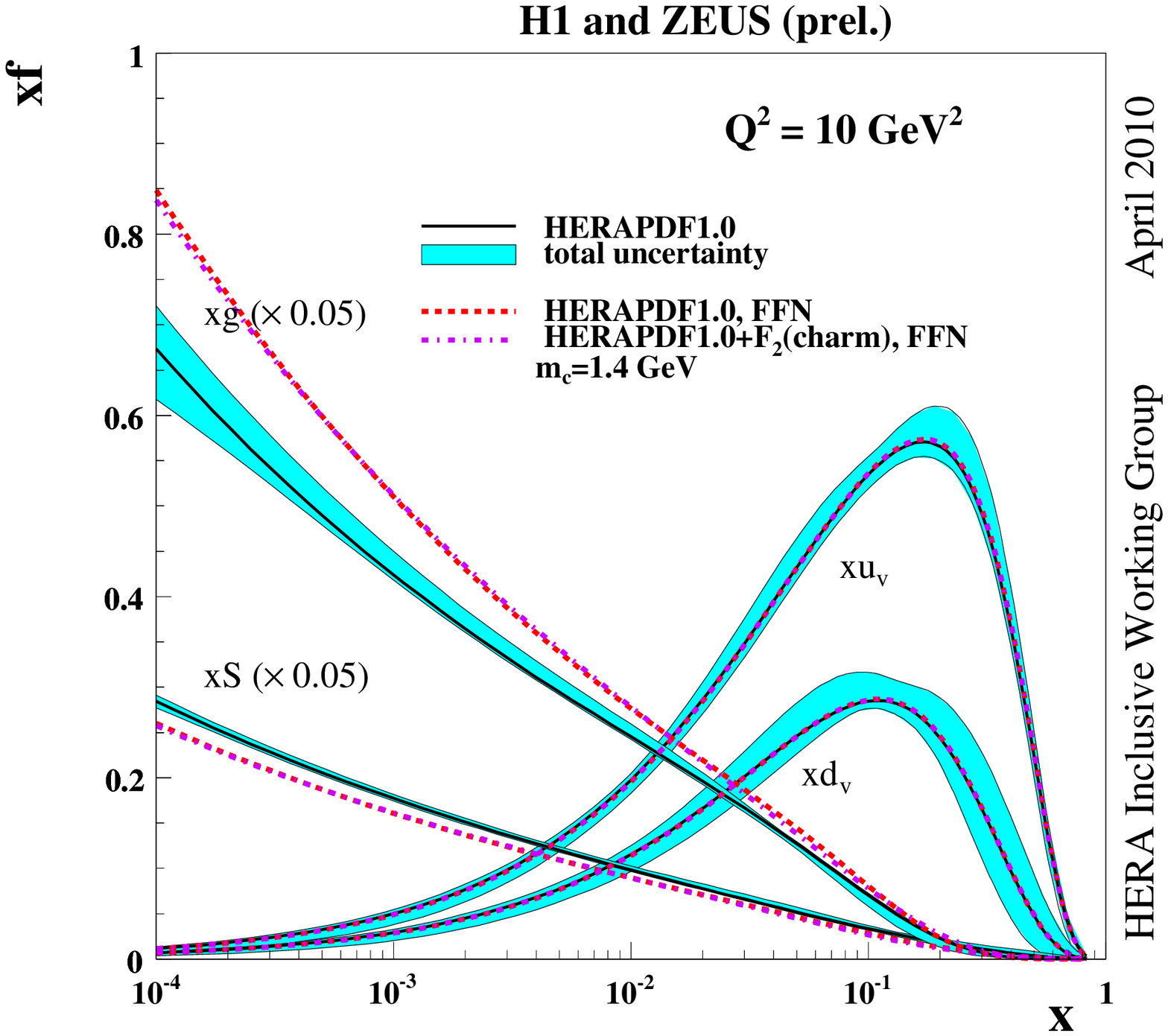,width=0.5\textwidth,height=6.0cm}}
\caption {Left: PDFs corresponding to the ACOT scheme fit for 
$m_c=1.4$ and $1.65~$GeV. 
Right: PDFs corresponding to the FFN scheme fit for 
$m_c=1.4$ and $1.65~$GeV.  }
\label{fig:acotffn}
\end{figure}

A different general-mass VFN
scheme is the ACOT scheme~\cite{acot}. 
In fits using the ACOT scheme the charm mass, $m_c=1.65~$GeV, is preferred, 
see Table~\ref{tab:chisq}. The fits to the data are similar to the 
RTVFN standard scheme fits. The corresponding PDFs are shown in 
Fig.~\ref{fig:acotffn}. The PDFs including the charm data for all 
these general-mass VFN 
schemes are not very different from the HERAPDF1.0 PDFs.

A Fixed Flavour Number (FFN) scheme can also be
used~\cite{ffn}. However, the CC $e^+$ and $e^-$ scattering data  
cannot be used in such a scheme since there is no complete calculation of 
the appropriate NLO coefficient functions. In the HERAPDF1.0 analysis
the CC data were used to constrain the valence quark PDFs. Thus for the FFN fits the valence quark parameters are fixed at 
their HERAPDF1.0 values and only the sea and gluon parameters are varied while 
fitting to NC data (524 points) and the charm data. 
Two further modifications are made for the 
FFN fit: firstly the heavy quark factorisation scale is chosen to be 
$Q^2 +4m_c^2$ (instead of $Q^2$ as for the RTVFN and ACOT schemes); secondly 
the running of $\alpha_s(Q^2)$ with $Q^2$ is calculated using only 3-flavours
(instead of using 3-, 4- and 5-flavour evolution with  matching 
prescription at flavour threholds as for the general mass VFN schemes). This 
means that a low equivalent value of $\alpha_s(M_Z)= 0.105$ must be used to 
ensure that $\alpha_s$ is not too high to give a good description of low-scale 
data. 
For these FFN fits charm is suppressed at threshold relative to the RT-VFN 
fits and  the value $m_c=1.4~$GeV is preferred by the data, see 
Table~\ref{tab:chisq}. The fits to the data are similar to those of the 
optimized TR-VFN scheme but the corresponding gluon PDF is strikingly 
different, see Fig.~\ref{fig:acotffn}. 
The FFN gluon is very much enhanced compared to that of the GMVFN schemes
(note the Sea only appears smaller because of the lack of a charmed parton).
Variations of the FFN fit have also been tried: $Q^2$ has been used as the 
heavy quark factorisation scale; a cut $Q^2 < 3000~$GeV$^2$ has been applied 
since the FFN scheme does not resum $ln(Q^2/m_c^2)$ terms; alternative PDF 
parametrisations have been tried. 
None of these modifications change the resulting
PDF shapes significantly, although the addition of an extra gluon parameter 
gives a modest improvement in $\chi^2$.

\begin{figure}[tbp]
%\vspace{-1.0cm} 
%\vspace*{5pt}
\centerline{
\epsfig{figure=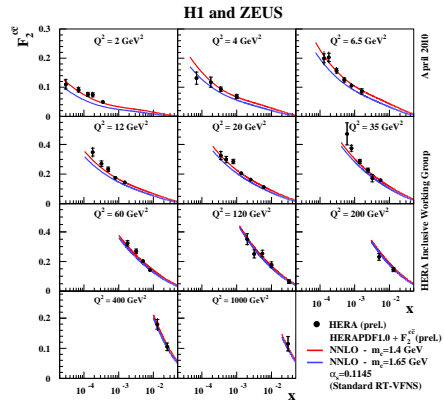,height=6.0cm}}
\caption {Comparison of $F_2$(charm data) to a PDF fit which includes 
these data for $m_c=1.4$ and $1.65~$GeV, using the standard RT-VFN scheme at 
NNLO and $\alpha_s=0.1145$.}
\label{fig:nnlo}
\end{figure}

Finally NNLO fits in the RT-VFN scheme were tried. Although Thorne has 
suggested possible variations in the heavy quark scheme at NNLO the 
differences between these schemes are far less than at NLO. The 
Standard NNLO scheme was used.
Fits were made for two different values of $\alpha_s(M_Z)$; $0.1176$ 
-the HERAPDF1.0 central value- and $0.1145$. 
This is because PDF fits with $\alpha_s(M_Z)$ free prefer the latter value 
at NNLO- albeit with a large error.
%$\alpha_s(M_Z)=1145\pm0.0042$
By contrast, at NLO the preferred value, 
$0.1166$, is much closer to the standard value.
The NNLO fits prefer $m_c=1.4~$GeV, whichever value of $\alpha_s(M_Z)$ is used,
see Table~\ref{tab:chisq}. Fig.~\ref{fig:nnlo} shows the NNLO fits to data for
$\alpha_s(M_Z)=0.1145$.
The shape of NNLO fit is somewhat different from the NLO fits 
and even though data at $Q^2=2~$GeV$^2$ are not included in the fit they are 
well described.

\section{Summary and Discussion}
 PDF fits have been made to the combined HERA-I inclusive NC and 
CC data and to the combined $F_2$ charm data. The charm data are sensitive to 
the value of the charm mass and the choice of heavy 
quark mass scheme. This has consequences for predictions of
the $W$ and $Z$ cross-sections at the LHC.  A larger charm mass means 
suppressed charm at threshold and thus the lighter quarks are enhanced 
to compensate. This results in a 
$2.5\%$ higher $W,Z$ cross-section for $m_c=1.65~$GeV as compared to 
$m_c=1.40~$GeV~\cite{mywz}.
Thus there is some uncertainty in these predictions resulting from the use of 
different heavy quark schemes and masses, which will be the subject of 
further study.

\end{document}